%% file: Main_paper.tex
\documentclass[lettersize,journal]{IEEEtran}
\IEEEoverridecommandlockouts
\usepackage{algorithm,algpseudocode} 
\usepackage{cite}
\usepackage[utf8x]{inputenc} 
\usepackage{amsmath}
\usepackage{bm}
\usepackage{amssymb,amsmath}
\usepackage[mathscr]{euscript}
\usepackage{graphicx}
\usepackage{epsfig}
\usepackage{tcolorbox}
\usepackage{grffile}
\usepackage{balance,color}
\usepackage{xcolor, soul}
\usepackage{multirow}
\usepackage{adjustbox} 
\usepackage{booktabs}
\usepackage{array}
\usepackage{adjustbox}
\sethlcolor{cyan}
\usepackage{comment}

\renewcommand{\textcolor}[2]{#2} 
\providecommand{\keywords}[1]{{\textit{Index Terms}}}
\newcommand{\fref}[1]{Fig.\ref{#1}}

\newcolumntype{C}[1]{>{\centering\arraybackslash}m{#1}}
\newcolumntype{M}[1]{>{\centering\arraybackslash}m{#1}}
\newcommand{\amrbox}[1]{
 \hspace{-1cm}\vspace{-2cm}\raisebox{2cm}{\begin{tabular}{p{5cm}}
 #1
  \end{tabular}} \\ 
}

\begin{document}

\title{\textcolor{red}{Learn to Slice, Slice to Learn : Unveiling Online Optimization and Reinforcement Learning for Slicing AI Services}} 

\author{Amr Abo-eleneen$^1$, Menna Helmy$^2$, Alaa Awad Abdellatif$^2$, Aiman Erbad$^2$, Amr Mohamed$^2$, Mohamed Abdallah$^1$ \\
\begin{tabular}{c}
	$^1$ College of Science and Engineering, Hamad Bin Khalifa University, Qatar \\ 	
  $^2$ College of Engineering, Qatar University, Qatar \\  
		Email: \{a.aboeleneen, alaa.abdellatif, moabdallah, aerbad\}@ieee.org,  \{menna.helmy,  amrm\}@qu.edu.qa,
		\thanks { This work was made possible by NPRP grant \#  NPRP13S-0205-200265 from the Qatar National Research Fund (a member of Qatar Foundation). The work of Amr Abo-eleneen was supported by GSRA grant \# GSRA9-L-1-0518-22022. The findings achieved herein are solely the responsibility of the authors.  }   
\end{tabular}
}   
\maketitle

\begin{abstract}
\textcolor{red}{ 
In the face of increasing demand for zero-touch networks to automate network management and operations, two pivotal concepts have emerged: “Learn to Slice” (L2S) and “Slice to Learn” (S2L). L2S involves leveraging Artificial intelligence (AI) techniques to optimize network slicing for general services, while S2L centers on tailoring network slices to meet the specific needs of various AI services. The complexity of optimizing and automating S2L surpasses that of L2S due to intricate AI services' requirements, such as handling uncontrollable parameters, learning in adversarial conditions, and achieving long-term performance goals. This paper aims to automate and optimize S2L by integrating the two concepts of L2S and S2L by using an intelligent slicing agent to solve S2L. Indeed, we choose two candidate slicing agents, namely the Exploration and Exploitation (EXP3) and Deep Q-Network (DQN) from the Online Convex Optimization (OCO) and Deep Reinforcement Learning (DRL) frameworks, and compare them. Our evaluation involves a series of carefully designed experiments that offer valuable insights into the strengths and limitations of EXP3 and DQN in slicing for AI services, thereby contributing to the advancement of zero-touch network capabilities.}

\end{abstract}

\section{Introduction\label{sec:Introduction}}


Network slicing (NS) has emerged as a pivotal tool in wireless networks, providing a mechanism to configure networks specifically for diverse applications and services. One groundbreaking application within this domain is integrating Artificial Intelligence (AI) technologies for automating the management and orchestration of the network's computational and network resources through NS, paving the way for the development of zero-touch networks to serve many services in parallel and on the same physical infrastructure.
This allows for the emergence of specialized research areas termed \textit{Learn to Slice} (L2S) and \textit{Slice to Learn} (S2L). The former aims to use AI to optimize network slices for varying general network services (e.g., remote monitoring). On the other hand, S2L focuses on the allocation of network slices and AI hyperparameters (e.g., number of epochs) to varying AI services for optimal overall results (e.g., highest overall accuracy or F1-score) without compromising their performance or reliability. S2L is beneficial in the case of AI task offloading when multiple devices or vehicles with limited resources offload their diverse machine learning tasks to nearby edge networks for training.

Unlike L2S, S2L presents a myriad of challenges specific to AI models that demand innovative solutions. These challenges come from the fact that AI is data and model-specific. Some of these complexities may include the ability of the slicing agent to :
\begin{itemize}
    \item Find the optimal choice of resources and AI hyperparameters for different models with different architectures (e.g., ANN, CNN, and GPT), complexity, and goals (e.g., classification) with minimal overloading to other services.
    \item \textcolor{red}{Consider the existence of multiple uncontrollable parameters that affect the accuracy of the models and their generalization, such as data quality (e.g., being independent and identically distributed (i.i.d), balanced, unique, and complete.., etc.).}
    \item Adapt to different models and data requirements.
    \item Learn within potentially untrusted environments (e.g., adversarial attacks by providing misleading data quality to the slicing agent to acquire more resources).
    \item Adhering to long-term goals (e.g., reducing total cost or preserving components' reliability in the long run).
    \item Take into account the stochastic and diverse nature of calculating the models' accuracy, as each model's accuracy is only known after applying a specific AI hyperparameter with certain data quality.
\end{itemize} Addressing these challenges is crucial for successfully implementing and automating the creation of perfect slicing solutions for AI. 

To tackle most of these formidable challenges and harness the potential of S2L, this paper mainly focuses on integrating intelligent slicing agents from L2S into S2L, effectively incorporating AI into the S2L process. Specifically, we explore two different slicing agents from two renowned learning frameworks: Online Convex Optimization (OCO) and Deep Reinforcement Learning (DRL). These frameworks are aptly presented by the Exponential-weight algorithm for Exploration and Exploitation 3 (EXP3) and the Deep Q-Network (DQN) algorithms. Their inherent characteristics make them suitable for supporting NS performance, especially catering to various AI services, as in the case of S2L. 
DQN, a key DRL framework, an agent interacts with its environment by observing states and selecting actions, receiving rewards based on action effectiveness. It uses neural networks to optimize cumulative rewards by refining action-state associations. Typically pre-trained in a specific environment, DQN can also adapt its training to real-world testing for added flexibility.
Instead, the EXP3, rooted in the OCO framework, is inherently an online learning algorithm. Without needing prior experience in the testing phase, EXP3 dynamically selects actions. Based on the rewards received, it adjusts the probability of specific actions, increasing their likelihood exponentially while ensuring a periodic exploration of new actions. 

Confronted with the aforementioned challenges, thoroughly examining such algorithms becomes imperative.
This paper seeks to thoroughly analyze the performance of the aforementioned frameworks in the context of the S2L problem. 
We dissect their strengths, limitations, and practical implications, aspiring to illuminate the pathway for optimized S2L.
Therefore, our contributions can be summarised as follows:
\begin{enumerate}
\item First, we discuss the efforts and challenges associated with S2L in 5G and beyond networks.
\textcolor{red}{\item Next, we formulate the problem of S2L, with the objective of maximizing the average accuracy of various AI services for classification tasks. This formulation focuses on the joint slicing of communication and computation resources at the Radio access network (RAN) and transport edge nodes while considering the diverse requirements of the different AI services related to computation, delay, and cost.} 
\item Then, we solve the formulated problem utilizing two of the most renowned algorithms, EXP3 and DQN. We demonstrate the effectiveness and limitations of these techniques in supporting S2L while adapting to varying environmental conditions such as sudden environmental change, timely constrained learning, adversary existence, and long-term goals.
\item Lastly, we conclude by highlighting some challenges and future research directions that are worth investigating.
\end{enumerate}

In what follows, Section \ref{sec:Sec2} discusses the key challenges associated with S2L and explores related works. Section \ref{sec:system} introduces the system architecture under consideration and outlines the formulated slicing optimization problem. Section \ref{sec:Benefits} presents our analysis and assessment of OCO and DRL frameworks in addressing the formulated problem. Finally, Section \ref{sec:conclusion} concludes the paper and explores potential directions for future research.

\begin{table*}[t]
\centering
\caption{\textcolor{black}{Summary of some of the related Work}}
\label{tab:related_work}
\begin{tabular}{|C{1.2cm}|C{2cm}|C{4cm}|C{3cm}|C{6cm}|}
\hline
\textbf{Ref} & \textbf{Application} & \textbf{Goal} & \textbf{Optimization method} & \textbf{Limitation} \\
\hline
\cite{Chen22} & S2L & Maximizing admitted training jobs in distributed machine learning (ML) through joint data collection and resource allocation. & Approximation via randomized rounding & 
\amrbox{
    \begin{itemize}
      \item No account for system dynamics or adversarial existence.
      \item Hyper-parameter tuning is not considered.
      \item No consideration of AI services KPIs.
      \item No consideration of long-term goals.
    \end{itemize}
  } 
\cline{1-4}
\cite{Edge-cloud22} & S2L & Cost minimization through efficient allocation of edge and cloud resources for distributed ML training jobs & Classical optimization & \\
\hline
\cite{Zeng22} & S2L & Energy minimization for federated edge learning (FEEL) through efficient radio resource management.   &  Convex optimization &
\amrbox{
    \begin{itemize}
      \item No account for system dynamics or adversarial existence.
      \item Considers single AI service and single access point (or edge server).
      \item Hyper-parameter tuning is not considered (for reference \cite{Zeng22}).
      \item No consideration of long-term goals.
    \end{itemize}}
\cline{1-4}
\cite{Lin21} & S2L & Minimize time and energy consumption through efficient AI service placement and resource allocation supporting model inference & Convex optimization &  \\
\hline
\cite{Bhardwaj2022} & S2L & Maximizing inference accuracy through the joint allocation of edge computing resources between training and inference tasks serving video analytics applications and selecting the best configurations for these tasks. & Classical optimization & \\
\cline{1-4}
\cite{Chen2023} & S2L & Accuracy maximization and delay minimization supporting AI training services through efficient task scheduling and resource allocation. & Non-dominated Sorting Genetic Algorithm & \amrbox{
\begin{adjustbox}{valign=c}
\begin{varwidth}{\linewidth}
    \begin{itemize}
      \item No account for system dynamics (for reference \cite{Chen2023}).
      \item Hyper-parameter tuning is not considered (for references \cite{Chen2023,Khani2023}).
       \item No account for adversarial existence.
      \item No consideration of long-term goals.
    \end{itemize}
    \end{varwidth}
\end{adjustbox}
  }
\cline{1-4}
\cite{Khani2023} & S2L & Efficiently allocating computing resources to maximize the accuracy improvement of model retraining tasks while incorporating the reuse of previous models to avoid redundant re-training. & Classical optimization & \\
\hline
{This article} & {S2L }&{Maximize the accuracy of multiple competing AI services + Assessing the benefits and limitations between DQN \& EXP3 as candidate solutions.} & AI techniques:  {DRL \& OCO}& 
\begin{itemize}
  \item {Multiple services with different KPIs.}
  \item {Learning in a dynamic environment.}
  \item {Learning under adversarial existence.}
  \item {Scalability is discussed.}
  \item {Considers long-term goals.}
\end{itemize}\\
\hline
\end{tabular}
\end{table*}

\section{\textcolor{red}{Related work \label{sec:Sec2}}}

In this section, we highlight the major challenges associated with slicing AI services and the related work of S2L. AI services fall under the Enhanced Mobile Broadband (eMBB) category, considering insights from recent relevant literature.
 
\paragraph{S2L \label{sec:why}}
\textcolor{red}{
AI services have multiple stages that enable continuous training and deployment of the underlying AI models. These stages include: 
1) Data collection and processing, 2) Model training, and 3) Model inference. Effective management of these stages and efficient allocation of the needed resources are vital for AI services to ensure service quality, the coexistence of multiple services, and seamless adaptation to changing network and user requirements. Hereafter, we outline some challenges associated with AI services:} 

\textbf{Limited resource-availability at the edge: }
AI services include computationally intensive processes such as model training and latency-sensitive tasks such as model inference. Due to the limited resources at the edge, efficient resource allocation mechanisms are required to fulfill the Quality of Service (QoS) and Quality of Experience (QoE) requirements per AI service while guaranteeing the coexistence and function of diverse services and virtual network functions. 

\textcolor{red}{\textbf{Distinct performance indicators of AI Services: } 
AI services have unique performance metrics that need careful evaluation alongside conventional indicators such as throughput. Key performance indicators (KPIs) may include \cite{AI21}: 
\begin{enumerate}
    \item {Learning speed} that measures the rate of fully training an AI model. This is limited by allocated computing resources and model hyperparameter tuning (e.g., data size, number of epochs, and learning rate). 
    \item {Learning accuracy} that defines the prediction results on input data in comparison to actual values. Statistical properties of acquired data and training epochs significantly impact accuracy, necessitating hyper-parameter tuning for desired performance.  
\end{enumerate}}

\textbf{User and Network dynamics: }
Resource allocation and hyper-parameter tuning for AI services must be performed while considering users and network dynamics, which may exhibit non-stationary changes throughout the learning time. Adapting to the underlying statistical patterns of the environment is crucial for achieving stationary performance. Furthermore, detecting and smoothly adjusting to shifts in the environment and data distribution, called concept drift, is paramount to preserving the model's performance over timely changes.

\textbf{Heterogeneity of AI models and learning hierarchies: } 
Efficiently allocating resources for AI services becomes complex due to the diverse architectures and hierarchies of underlying AI models, such as centralized versus distributed structures \cite{AI21,Edge-cloud22}. Addressing considerations like whether end-users perform local computation (for training or inference) \cite{Zeng22,Lin21}, whether user data is offloaded to the edge \cite{Chen22,Edge-cloud22}, and determining the optimal utilization of cloud resources \cite{Edge-cloud22} all require careful design of NS schemes to ensure efficiency in AI service delivery.

It is important to note that one S2L architecture was recently outlined in \cite{Wu2022Feb}, focusing only on assigning pre-created AI training parameters (e.g., algorithm, training manner) to the different AI models while satisfying environment constraints. Conversely, our proposed solution maximizes the joint training accuracy by assigning both model AI hyperparameters, computation, network, and data percentage to train the AI models optimally.

\textcolor{red}{
In Table~\ref{tab:related_work}, we summarize the goals and limitations of some of the key relevant works related to S2L. We also highlight the distinctive aspects of our paper compared to these works. 
While the methods presented to address the S2L problem encompass a range of techniques, including classical methods, convex optimization, and approximation algorithms, they offer only partial solutions to the complex challenges of optimizing AI services. These methods may not fully align with the zero-touch network approach, as they can struggle to rapidly adapt to the dynamic and continuous changes occurring in the RAN environment.
Moreover, many works in Table~\ref{tab:related_work} fail to account for real-world deployment scenarios. In contrast, this article integrates L2S techniques, such as DRL, to tackle the S2L problem, emphasizing intelligent resource slicing for AI services and effective management of challenging environmental conditions.}\\\indent 
In what follows, we describe the problem with its requirements and constraints. Then, we explore two popular frameworks of approaches (i.e., OCO and DRL) designed to facilitate real-time decision-making in diverse environmental conditions and compare their respective advantages and limitations. Based on our knowledge, this is the first article to discuss this.

 

\section{S2L Architecture, Scenario and Problem Formulation \label{sec:system}}
\textcolor{red}{In this section, we describe the considered system architecture along with the scenario and formulated optimization problem for S2L. }
\subsection{System Architecture\label{sec:Architecture}} 

\begin{figure*}
	\centering
		\scalebox{0.75}{\includegraphics{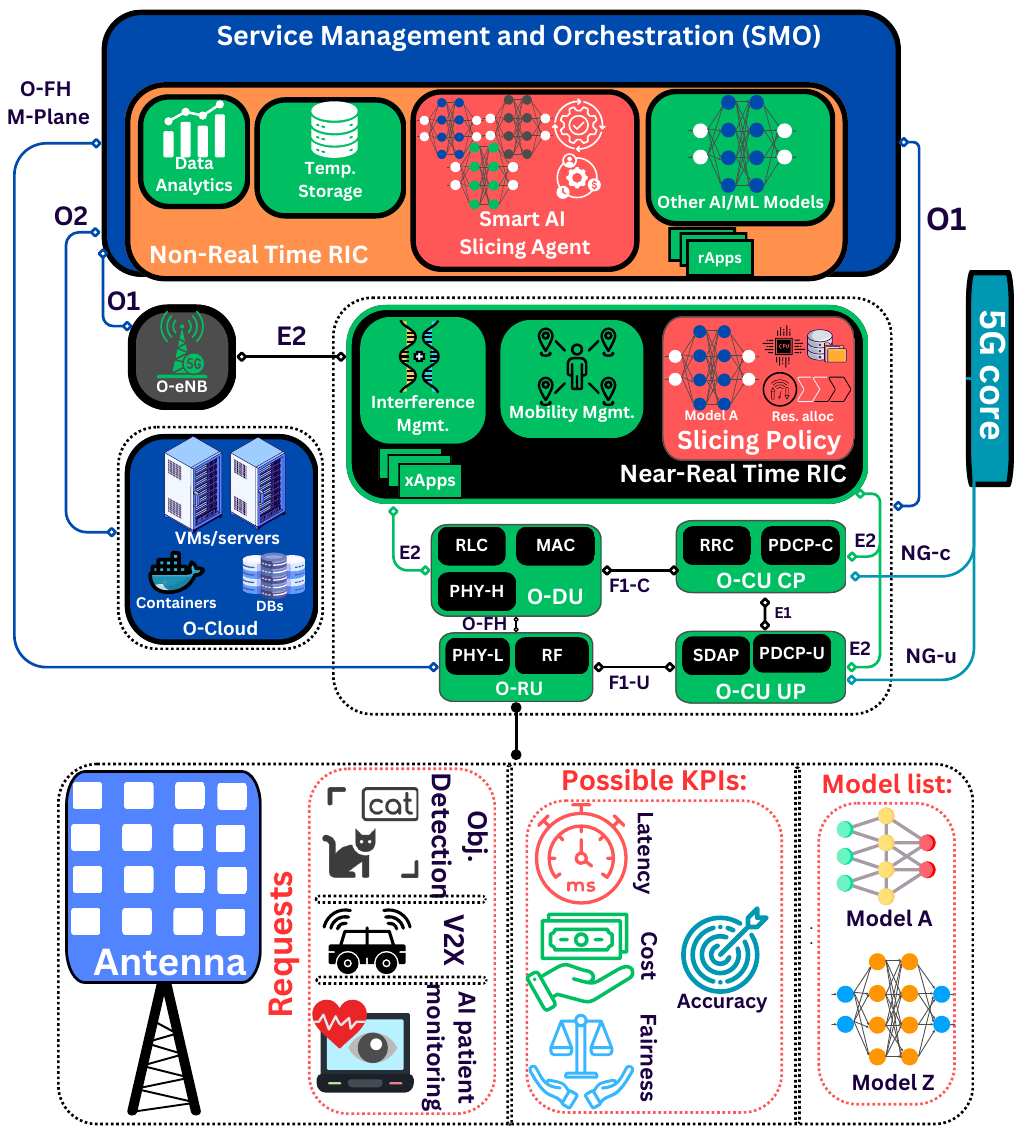}}
	\caption{\textcolor{red}{S2L architecture based on latest O-RAN architecture \cite{oranarch}} }
	\label{fig:systek_model}
\end{figure*}

\textcolor{red}{The architecture presented in \fref{fig:systek_model} illustrates the overall design of our S2L system integrated within the Open RAN (O-RAN) framework \cite{ORAN,oranarch}. The figure overviews various O-RAN components, their communication interfaces (e.g., O1, O2, E2, O-FH), and their responsibilities. Firstly, the Service Management and Orchestration (SMO) component and its subcomponent Non-Real Time RAN Intelligent Controller (RIC) are responsible for orchestrating end-to-end services' slices and optimization of different RAN elements through the use of data analytics, temporary storage, and intelligent automation apps (rApps). Secondly, the Near-Real Time RIC optimizes RAN performance in near-real time by making intelligent decisions such as dynamic resource allocation and handover management based on trained machine learning models. Thirdly, Open-Cloud (O-Cloud) offers a virtualized and containerized infrastructure for deploying and managing O-RAN functions, ensuring scalability, reliability, and flexibility for the RAN components. The Open enhanced evolved NodeB (O-eNB) serves as the base station that provides connectivity to user devices and facilitates communication between the user equipment and the core network.
Additionally, the architecture includes the Open virtualized RAN components such as the Open Distributed Unit (O-DU), Open Control Unit (O-CU), and Open Radio Unit (O-RU). 
In this architecture, our system starts by collecting the AI services' training requests from the O-RU. Each request includes the AI training task (e.g., classification) along with the user-selected architecture from a limited list of available models, the size of the dataset, and the needed KPIs.
After that, our AI slicing agent, highlighted in red, works in the background within the SMO's Non-Real Time component for training the agent (e.g., gathering data and training the slicing agent, such as the DQN agent). Once trained, it is deployed in the Near-Real-Time RIC to make quick, adaptive decisions to jointly select the optimal virtual resources and AI hyperparameters per each AI service's model to achieve the highest average accuracy. 
}

\textcolor{red}{
\subsection{Scenario}
To create a realistic scenario, we assume the presence of multiple users, each aiming to train their AI model through a central entity. The objective of our agent is to train these models with the highest possible average accuracy by joint slicing of communication and computation at the RAN and edge computing devices. This scenario mirrors real-world AI training situations, such as smart vehicles offloading their AI model training to roadside units or in Edge AI, where users or drones offload model training to a central entity.
The training process begins when users send their data sizes and specify their desired AI model architecture (e.g., a CNN-LSTM model) from a predefined set of options. The smart agent then manages these requests in two stages:
\begin{enumerate}
    \item \textbf{Stage One}: The agent allocates the necessary bandwidth for each user's AI service, allowing them to transfer their training data to the RAN. It also determines the proportion of the user's data to be used for training (e.g., 70\%).
    \item \textbf{Stage Two}: The agent then selects appropriate computing resources (e.g., CPU frequency allocation) and AI hyperparameters, such as the learning rate and number of training epochs, for the models running on the edge nodes.
\end{enumerate}
In our scenario, we adopt three different AI services; each is a variant of a CNN-LSTM Deep Learning classification model for physical health and activity analysis and is based on a public dataset \cite{xxx}. The three models have different KPI constraints for learning latency and learning accuracy, and the goal of our intelligent slicing agent is to optimize resource allocation to meet the KPIs while maximizing the average accuracy of the AI models and adhering to different constraints, as will be described later.  
}

\subsection{Problem Formulation \label{sec:Cases}}   
\textcolor{black}{Our problem, denoted as \textbf{P}, aims to maximize the utility function $U$ across all timesteps $\forall{t} \in \mathcal{T}$ for all AI models $ \forall{m} \in \mathcal{M}$ i.e., $U^{(t)} = \sum_{m = 1}^{M}{A_{m}(N_{m}^{(t)},k_{m}^{(t)}) \times q_{m}^{(t)}}$. In this utility function, $A_m$ denotes the accuracy of model $m$ (a function of data length $N_{m}^{(t)}$ and a number of epochs $k_{m}^{(t)}$), and $q_{m}^{(t)}$ symbolizes the uncontrollable data quality received from the users provided a dataset for model $m$.\\
The decision variables used to solve problem \textbf{P} are the parameters that define a slice for each AI model: 
\begin{itemize}
    \item The fraction of dataset sizes gathered from users.
    \item Data rates for transferring datasets from users to edge nodes.
    \item Edge CPU frequency for model training.
    \item The number of training epochs.
\end{itemize}
Moreover, a solution to problem \textbf{P} must consider various constraints, including computational limits (i.e., edge capacity), communication limits (i.e., RAN bandwidth), latency requirements (i.e., training delay and network delay), and training cost/budget.}
\textcolor{red}{We also note that $q_{m}^{(t)}$ was used to simulate one difficulty in training AI models. In our case, it is a percentage representing the data quality of the training dataset for a given model at a time slot. While there is no standard model in the literature for measuring $q_{m}^{(t)}$, it can be assessed based on factors such as whether the data is i.i.d., balanced, complete, accurate, and unique. Moreover, $q_{m}^{(t)}$ can dynamically change over time according to a stochastic process (i.e., according to a stationary distribution) or according to a non-stationary distribution imposed by an attacker (i.e., when an attacker wants to mislead DRL or OCO for knowing optimal decision variables for training). Consequently, it is crucial to adopt an intelligent solution to learn the optimal allocation of resources and tune the learning models' hyper-parameters under changing system dynamics (i.e., quality factor) and different models' constraints. Only in subsection \ref{ssub:work} the quality factor will be considered to model adversarial attackers that intend to trick DRL and OCO frameworks. 
} 
 

Tackling problem \textbf{P} is challenging due to the constraints of limited edge resources, coupled with stringent latency and cost limitations of training different models. We note here that the learning latency (or processing delay) is estimated as a function of the acquired dataset size and model training delay (which is a function of the number of epochs and CPU frequency) \cite{Pervasive22}. The communication delay is based on factors such as transmitted data size, transmission rate, and channel access delay, as defined in \cite{awad2011energy}. \textcolor{black}{The learning accuracy is estimated as a function of dataset length and the number of training epochs acquired using a regression model on a certain dataset.} \textcolor{black}{We remark that the accuracy function is only known through trial and error; therefore, classical convex optimization is not applicable.}
Additionally, given that more epochs and larger data sizes often improve model accuracy, it is crucial to balance the trade-off between obtained accuracy, cost, and latency while accounting for the available computational and communication resources. 




\section{Strengths and Limitations  of OCO and DRL \label{sec:Benefits}}
\input{DRLvsOCO}





\section{Conclusions and Future Directions\label{sec:conclusion}}

\textcolor{red}{This paper investigates the use of intelligent slicing agents to automate and optimize the S2L problem through joint slicing of communication and computation resources across the RAN and edge nodes. We leverage and compare EXP3 and DQN from OCO and DRL frameworks to solve this problem while supporting diverse AI services and their challenges. Specifically, both algorithms have undergone evaluations in various scenarios to demonstrate their learning efficiency, scalability, resilience in the presence of adversaries, and alignment with long-term objectives to provide the reader with each solution's advantages and limitations.  The presented work opens up several promising research directions for future exploration, including:} 

\subsubsection{Drift detection}
Our findings show that drift in the environment may result in system performance degradation for some time until the decision-making agent can adapt to the changes. Hence, It is crucial to establish mechanisms that can swiftly identify when the performance of the decision-making agent begins to decline or deviate from expected standards. Identifying such a deterioration can be a vital precursor in initiating a retraining process to ensure the reliability of decision-making.     


\subsubsection{S2L with Meta-learning}
\textcolor{black}{Since S2L agents encounter various models, it would be beneficial for the slicing agent to utilize the knowledge gained from previous models and apply it to new models without extensive retraining. A promising research area to address this issue is meta-learning. Meta-learning leverages knowledge from previously encountered tasks to enable quick adaptation to new tasks with minimal system interactions. Therefore, integrating meta-learning with DRL or OCO frameworks to S2L deserves further investigation.}

\subsubsection{Joint allocation for training and inference tasks} 
In our work and the studies \cite{Zeng22,Chen22,Edge-cloud22}, resource allocation was conducted with a focus on training tasks only. Conversely, inference tasks were only considered in the work presented in \cite{Lin21}. However, no solution was presented to allocate resources for training and inference tasks simultaneously. Therefore, there is a need for innovative solutions that tackle both aspects simultaneously. In that sense, distributed inference based on a cloud of IoT devices' resources can be used as part of the slice.
\subsubsection{ Varying action space} 
In many research studies, DQN and EXP3 have often been applied with fixed action spaces. This means that the set of available actions remains constant across different states. However, this approach may not accurately represent real-world scenarios, especially in environments where system scalability plays a role. In such dynamic environments, where the system can scale up, e.g.,  by increasing available resources or adding more virtual machines (VMs),  the set of available actions can change significantly and expand. Thus, developing more adaptive systems with variable action spaces can be crucial in such scenarios.  

\balance 

\bibliographystyle{IEEEtran}
\bibliography{ref}

\end{document}

%% file: DRLvsOCO.tex
Given the requirements and challenges of S2L discussed in Section \ref{sec:Sec2}, we now highlight the effectiveness and limitations of EXP3 and DQN in fulfilling these requirements through an array of experiments that assess their capabilities in terms of: 1) Converging and adapting to new environments, 2) Learning under the existence of an adversary (Adversarial resilience), 3) Applicability in real-time scenarios (Learning/Execution cost), and 4) Their adherence to long-term goals, such as preserving the reliability of a hardware component for the longest time possible.
\subsection{Comparison Considerations}
\textcolor{black}{Since DQN and EXP3 originate from distinct frameworks, we needed to have a fair comparison between them, and this entails the careful choice of 1. states, 2. actions, and 3. algorithm improvements.
Firstly, since DQN considers stateful systems as opposed to EXP3, we needed to align DQN more closely with EXP3. Hence, for the experiments detailed in sections \ref{ssub:conv}, \ref{ssub:exec}, and \ref{ssub:work}, we adopted a single-state environment for DQN. This decision aimed to mirror a scenario in the initial experiment where the single state mimics minimal context, akin to EXP3. Such a setup simplifies the DQN agent's learning process, as its objective involves selecting optimal actions throughout the episode, aligning with our specific training goal of optimizing action selection. However, when an agent's actions impact its state (e.g., budget) or state interdependencies exist, a single-state approach becomes inadequate, prompting a multi-state environment. Accordingly, the latest experiment described in section \ref{ssub:adh} embraced a multi-state environment.\\
Secondly, concerning the action space for DQN and EXP3, both shared an identical structure, encompassing a list of actions. Each action constituted a list of vectors, where a vector defines computational slice resources per model. A sample vector includes data size, number of training epochs, computing resources, and data rate for one AI service. Furthermore, the reward function for both algorithms was tailored to our optimization objective: an agent earns a higher reward for increased average accuracy. If any constraint is unmet, the agent receives a negative reward in DQN and a zero reward in EXP3.}

Finally, we considered the VANILLA versions of both algorithms exclusively. This approach eliminates the influence of introduced enhancements. Our rationale is to create a baseline comparison highlighting the core performance distinctions between the two algorithms. Utilizing unaltered VANILLA versions enables us to accurately assess each algorithm's inherent capabilities and limitations, unaffected by additional modifications. 

To ensure consistency in results, we used the same hardware across experiments: CPU - Intel® Core™ i7-9700 Processor, GPU - RTX 2080 TI, and 32GB RAM. DQN and EXP3 parameters were chosen based on the closest results. DQN parameters included batch size 2048, gamma 0.90 (initial three experiments) and 0.92 (final experiment), buffer size 1M, epsilon decay 0.99, and final exploration epsilon of 0.05. EXP3 employed an exploration rate of 0.001. \textcolor{red}{Additionally, to simulate the accuracy behavior per each AI model, we empirically derive a three-term exponential model with six different coefficients considering a custom CNN-LSTM Deep Learning classfication model for physical health and activity analysis based on a public dataset. We alter the coefficients to reproduce three distinct AI models that vary in terms of the relationship between the model parameters (e.g., number of epochs and data size) and the corresponding accuracy. The simulated environment was created using the Gymnasium library, and both implementations of DQN and EXP3 were created using Python.}
In what follows, we explain the experiments we carried out to assess DQN and EXP3 solutions. 
\begin{figure*}[t]
    \centering    \includegraphics[scale=0.95]{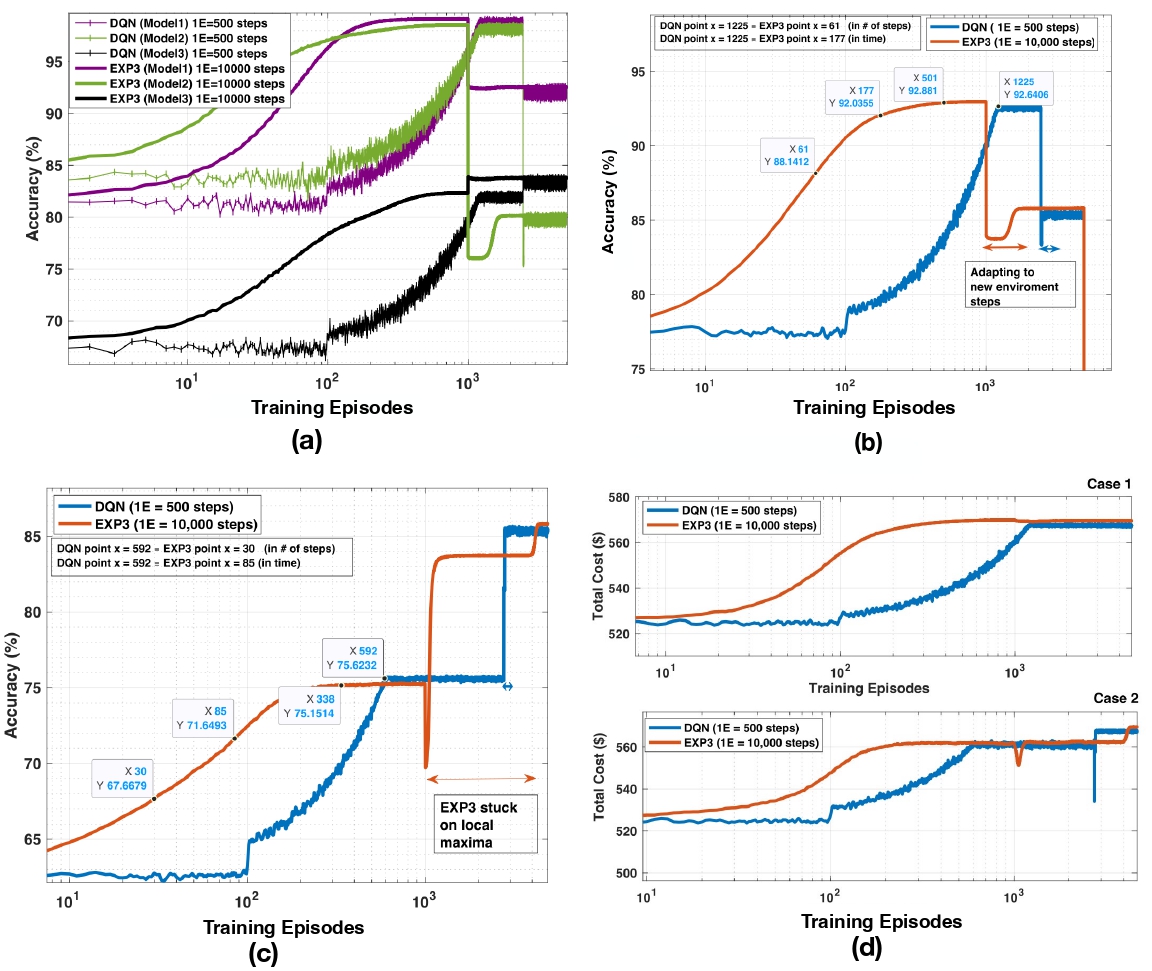}
    \caption{\textcolor{red}{Convergence and adaptation performance of DQN and EXP3: (a) shows \textit{Case 1} convergence and adaption of the different models, (b) and (c) shows average accuracy for both algorithms along with steps equivalence for \textit{Case 1} and \textit{Case 2} respectively and (d) shows incurred costs for both cases.}}
    \label{fig:conceptdriftfig}
\end{figure*}
 \subsection{Experiment 1: Convergence and Adaptation}
 \label{ssub:conv}
   Considering that environments frequently undergo dynamic changes, this experiment explores the capability of both the EXP3 and DQN algorithms to handle changes in such dynamic settings. Our approach involves training both agents to fine-tune computational slices for three distinct AI initial models, each with unique requirements and constraints.
   As we achieve convergence with these initial models, we introduce three unexplored models. Subsequently, we track the number of steps and time both approaches take to adjust. This experiment highlights how each approach utilizes its accumulated knowledge to adapt to the changing environmental dynamics.

   To comprehensively evaluate the effectiveness of the adaptation process, we replicated this experiment for two cases. The \textit{first case} focuses on the convergence of two sets of closely related models (i.e., initial and unexplored models are close in terms of model parameters and corresponding accuracy), providing insights into how the approaches handle minor disparities. In the \textit{second case}, we consider two sets of substantially distinct models, shedding light on the approaches' performance in accommodating more significant differences.

     \textcolor{black}{The results in \fref{fig:conceptdriftfig} show accuracy differences between DQN and EXP3. Generally, EXP3 has smoother and more stable learning than DQN, likely due to their different exploration, exploitation, and update mechanisms. DQN's epsilon-greedy strategy introduces fluctuations by periodically exploring random actions and updating its policy after multiple episodes. In contrast, EXP3 uses a probabilistic approach, exploring all actions equally at first and then updating action weights based on rewards immediately, allowing for continuous, reward-based adaptation and smoother learning.}
   
We also observe that EXP3 training typically involves significantly more steps than DQN. Given this, we chose to compress its outcomes for improved clarity in visualization. Specifically, 1 episode is equivalent to 500 training steps for DQN and 10,000 training steps for EXP3.
   In \fref{fig:conceptdriftfig} (a), we show a sample of the convergence and adaptation of the first case, highlighting the accuracy per each initial model and the adaption to new models by EXP3 and DQN indicated by the sudden drop and re-convergence in the graph, on and beyond the $10^3$ training episode. 
   We extend the view of \fref{fig:conceptdriftfig} (a) in \fref{fig:conceptdriftfig} (b), where we show the average accuracy of the three models gained by both algorithms. The graph shows that DQN takes significantly fewer steps to converge on the best average accuracies for the initial models and adapt to the new models' requirements than EXP3 (indicated by the red and blue double-headed arrows). This adaptation might be due to the continuous exploration of the DQN agent along with the previously collected experience that swiftly tunes the agent's value function to a better policy. On the other hand, EXP3's small learning rate can cause the learning to stall for some time. Similar lessons can be learned from \fref{fig:conceptdriftfig} (c) for the second case scenario. \textcolor{red}{Finally, \fref{fig:conceptdriftfig} (d) illustrates the convergence of the total training cost for the ML models in both scenarios (cases 1 and 2). The training cost for an ML model is calculated as the sum of: (a) the number of computing resources allocated by the model multiplied by the unit price of computing resources, and (b) the data rate allocated by the model multiplied by the unit cost of data rate.}
   
   The key takeaway from this experiment is twofold: firstly, both algorithms demonstrate the capability to attain the optimal solution for our problem. Secondly, it is worth noting that DQN exhibits the capacity to learn and adapt with remarkable efficiency, requiring significantly fewer training steps compared to EXP3.


\subsection{Experiment 2: Real-time applicability}
     \label{ssub:exec}
    Considering the significance of learning time in real-time situations, we analyzed the execution duration of EXP3 and DQN methods. Our previous experiment in \fref{fig:conceptdriftfig} (a), (b), and (c) highlighted DQN's efficient convergence and adaptation. However, given the inherent differences between both algorithms, evaluating the time implications of these steps is crucial, as the time taken per step varies significantly between the two. 
    
  In another experiment, we compared the execution times of the EXP3 and DQN algorithms across various batch sizes (e.g., 512, 1024, and 2048). In our experiment, 5.5 million training steps for EXP3 were equivalent to 2 million for DQN when using a batch size of 2048. This equivalence gradually decreases with smaller batch sizes. Batch size significantly influences the balance between accuracy and speed in neural networks, as observed in DQN. Our findings indicate that for batch sizes of 512 and 2048, approximately 2.5 to 3 steps of EXP3 equate to a single step in DQN. This suggests that EXP3's steps carry a notably lighter computational load than DQN's.

It is worth noting that the typical implementations of EXP3 rely exclusively on the CPU for all processes. In contrast, DQN leverages both the CPU and GPU: the CPU manages environmental interactions, the replay buffer's storage and retrieval, and random exploration, while the GPU focuses on the training and fine-tuning of neural networks. In our study, both the CPU and GPU were employed for DQN to reflect the practical usage of the algorithm.

Furthermore, to gain a holistic understanding of which algorithm initially trains faster, we highlighted the convergence point of DQN and indicated the equivalent points based on time and steps as depicted in \fref{fig:conceptdriftfig} (b) and (c). For instance, in \fref{fig:conceptdriftfig} (b), DQN begins converging at episode 1225, corresponding to episode 61 in terms of steps and episode 177 in terms of time for EXP3. Meanwhile, EXP3 starts its convergence later in episode 501. Comparing the two sub-figures, DQN converges more efficiently in step count and duration.

In summary, EXP3's lightweight CPU-dependent steps make it more flexible for deployment, unlike DQN, which requires GPU and CPU resources. Additionally, DQN proved more efficient regarding step count and adaptation time.
    
   \subsection{Experiment 3: Working under Adversarial existence:}
    \label{ssub:work}
    Nowadays, AI systems are employed everywhere, opening the door for different attackers to attempt to disrupt these systems by some deliberate adversarial behavior. In this experiment, we assess how well each algorithm can learn in such an environment. For this scenario, we assume the existence of an adversary falsifying the data quality factor, which is a significant parameter for calculating the accuracy of the models along with the number of epochs and data size, as discussed earlier in \ref{sec:Cases}. The change in the data quality will cause the agent to learn different accuracies to the same data size and the number of epochs for each underlying AI model, which will eventually give the agents a mixed performance signal, causing some learning difficulties. Moreover, we considered two distinct cases of attackers for a comprehensive assessment. The first involves an infrequent attacker who changes the data quality value once every three steps, while the other entails a frequent attacker who does the same every other system step. Moreover, due to the potential impact of action space size on agent learning (e.g., smaller sets being easier to learn from), we opt to analyze two action spaces: a smaller one (4K actions) and a larger one (18K actions).

   The first four rows of Table \ref{tab:performance} offer a range of insightful observations. To start, EXP3 showcased superior proficiency compared to DQN in learning under adversarial existence. Additionally, the impact of the attacker's frequency varies based on the action space size (S indicates small and B indicates Big action space). Specifically, while low and high frequencies within a smaller space exhibit minimal differences (0-1\%), a substantial contrast emerges between scenarios with low and extreme adversarial presence in a larger action space (7-44\%). Furthermore, in the context of a larger action space and the introduction of an adversary, we observe a significant distinction between DQN and EXP3, as indicated by the Advantage-percentage (Adv.) column. The Advantage column represents the percentage difference between the accuracy of one solution (e.g., DQN) multiplied by the number of steps compared to the other solution (e.g., EXP3).

    From this experiment, we understand that in the presence of adversarial behavior and varying action space sizes, EXP3 consistently outperforms DQN in terms of learning efficiency.


\subsection{Experiment 4: Optimizing and adhering to long-term Goals:}
     \label{ssub:adh}
In prior experiments, the primary focus was on optimizing each model's computational slice without explicit consideration of long-term objectives. In this experiment, we aim to demonstrate how each algorithm performs when optimizing for a long-term goal, even when this goal is not explicitly expressed within a utility function.

We have chosen to focus on the long-term goal of preserving a critical hardware component of the system. Our system relies heavily on storage devices, such as HDDs, for two primary operations:
\begin{enumerate}
    \item \textit{Writing/Storing}: In this operation, our system acquires data from end devices via the network and writes it to an HDD.
    \item \textit{Reading}: In this operation, the system trains various machine learning models by iterating over the acquired data for multiple epochs during model training.
\end{enumerate}
\textcolor{black}{In this context, we have introduced the concept of a Read/Write budget, a shared allocation for reading and writing. The reduction of this budget is influenced by two factors: one determined by the chosen number of epochs (for read operation) and another based on the data size (for write operation) selected by the agent for each of the three models. Specifically, we have established these values as 0.2 and 0.5, respectively, to indicate that the cost of the write operation exceeds that of the read operation.}

    As shown in the final four rows of Table \ref{tab:performance}, the outcomes demonstrate the comparison between DQN and EXP3 as they operate within various reliability budgets. These findings showcase the distinct behaviors of each algorithm based on their given information without explicit guidance. DQN opted to increase the number of steps, allowing it to train more models, although this came at the expense of lower accuracy. In contrast, EXP3 maintained a higher accuracy level by training with fewer but more focused steps. This discrepancy could stem from DQN's advantage in recognizing the environmental context, enabling it to make sub-optimal decisions that lead to greater overall benefits.
    This performance difference is apparent in the Advantage percentage column, where DQN outperformed EXP3 by a range of 67\% to 141\% in terms of overall performance (steps multiplied by accuracy). This distinction is particularly pronounced in the scenario with the highest budget, where DQN managed to train about 2.7 times more models while only experiencing a 10\% drop in accuracy.

    From the above, we conclude that DQN can intelligently act inside the environment to adhere to a long-term goal through multiple iterations of trial and error with the help of being stateful, unlike EXP3.


\begin{table}[ht]
\caption{Comparison of DQN and EXP3 performances with budget and adversary information.}
    \label{tab:performance}
    \centering
    \adjustbox{max width=1\linewidth}{
    \begin{tabular}{ccccccccccc}
        \toprule
        \multirow{2}{*}{\shortstack{\textbf{Action}\\ \textbf{size}}} & \multirow{2}{*}{\shortstack{\textbf{R/W}\\ \textbf{Budget}}} & \multirow{2}{*}{\shortstack{\textbf{Adversary}\\ \textbf{Exist.}}} & \multicolumn{3}{c}{\textbf{DQN}} & \multicolumn{3}{c}{\textbf{EXP3}} \\
        \cmidrule(lr){4-6} \cmidrule(lr){7-9}
        & & & \textbf{Steps} & \textbf{Acc.} & \textbf{Adv. (\%)} & \textbf{Steps} & \textbf{Acc.} & \textbf{Adv. (\%)} \\
        \midrule
        S & N/A & Low & 500 & 98\% & 0 & 500 & 98\% & 0 \\
        S & N/A & Hi & 500 & 96\% & -1.0\% & 500 & 97\% & \textbf{+1.0\%}  \\
        B & N/A & Low & 500 & 80\% & -8.0\% & 500 & 87\% & \textbf{+8.8\%} \\
        B & N/A & Hi & 500 & 42\% & -51.2\% & 500 & 86\% & \textbf{+104.8\%} \\
        \midrule
        S & 500 & N/A & 103 & 69 & \textbf{+67.2\%} & 50 & 85 & -40.2\% \\
        S & 1000 & N/A & 213 & 71 & \textbf{+77.9\%} & 100 & 85 & -43.8\% \\
        S & 2000 & N/A & 411 & 75 & \textbf{+141.8\%} & 150 & 85 & -58.6\% \\
        \bottomrule
    \end{tabular}
    }
    
\end{table}

%% file: Main_paper.bbl
\begin{thebibliography}{10}
\providecommand{\url}[1]{#1}
\csname url@rmstyle\endcsname
\providecommand{\newblock}{\relax}
\providecommand{\bibinfo}[2]{#2}
\providecommand\BIBentrySTDinterwordspacing{\spaceskip=0pt\relax}
\providecommand\BIBentryALTinterwordstretchfactor{4}
\providecommand\BIBentryALTinterwordspacing{\spaceskip=\fontdimen2\font plus
\BIBentryALTinterwordstretchfactor\fontdimen3\font minus \fontdimen4\font\relax}
\providecommand\BIBforeignlanguage[2]{{%
\expandafter\ifx\csname l@#1\endcsname\relax
\typeout{** WARNING: IEEEtran.bst: No hyphenation pattern has been}%
\typeout{** loaded for the language `#1'. Using the pattern for}%
\typeout{** the default language instead.}%
\else
\language=\csname l@#1\endcsname
\fi
#2}}

\bibitem{Chen22}
M.~Chen, et~al., ``Joint data collection and resource allocation for distributed machine learning at the edge,'' \emph{IEEE Transactions on Mobile Computing}, vol.~21, no.~8, pp. 2876--2894, 2022.

\bibitem{Edge-cloud22}
I.~Sartzetakis, et~al., ``Resource allocation for distributed machine learning at the edge-cloud continuum,'' in \emph{ICC 2022 - IEEE International Conference on Communications}, 2022, pp. 5017--5022.

\bibitem{Zeng22}
Q.~Zeng, et~al., ``Energy-efficient radio resource allocation for federated edge learning,'' in \emph{2020 IEEE International Conference on Communications Workshops (ICC Workshops)}, 2020, pp. 1--6.

\bibitem{Lin21}
Z.~Lin, S.~Bi, and Y.-J.~A. Zhang, ``Optimizing ai service placement and resource allocation in mobile edge intelligence systems,'' \emph{IEEE Transactions on Wireless Communications}, vol.~20, no.~11, pp. 7257--7271, 2021.

\bibitem{Bhardwaj2022}
R.~Bhardwaj, et~al., ``{Ekya: Continuous learning of video analytics models on edge compute servers},'' in \emph{{19th USENIX Symposium on Networked Systems Design and Implementation (NSDI)}}.\hskip 1em plus 0.5em minus 0.4em\relax USENIX Association, 2022, pp. 119--135.

\bibitem{Chen2023}
T.~Chen, Q.~Tang, and G.~Liu, ``{Efficient task scheduling and resource allocation for AI training services in native AI wireless networks},'' \emph{{IEEE Network}}, vol.~37, no.~1, pp. 637--642, 2023.

\bibitem{Khani2023}
M.~Khani, et~al., ``Recl: Responsive resource-efficient continuous learning for video analytics,'' in \emph{20th USENIX Symposium on Networked Systems Design and Implementation (NSDI 23)}, 2023, pp. 917--932.

\bibitem{AI21}
M.~Li, et~al., ``Slicing-based artificial intelligence service provisioning on the network edge: Balancing ai service performance and resource consumption of data management,'' \emph{IEEE Vehicular Technology Magazine}, vol.~16, no.~4, pp. 16--26, 2021.

\bibitem{Wu2022Feb}
W.~Wu, et~al., ``Ai-native network slicing for 6g networks,'' \emph{IEEE Wireless Communications}, vol.~29, no.~1, pp. 96--103, 2022.

\bibitem{oranarch}
\BIBentryALTinterwordspacing
O.-R. ALLIANCE, ``{O-RAN Architecture Overview {\ifmmode---\else\textemdash\fi} oran master documentation},'' Sept. 2024, [Online; accessed 23. Sep. 2024]. [Online]. Available: \url{https://docs.o-ran-sc.org/en/latest/architecture/architecture.html}
\BIBentrySTDinterwordspacing

\bibitem{ORAN}
A.~Garcia-Saavedra and X.~Costa-Pérez, ``O-ran: Disrupting the virtualized ran ecosystem,'' \emph{IEEE Communications Standards Magazine}, vol.~5, no.~4, pp. 96--103, 2021.

\bibitem{xxx}
\BIBentryALTinterwordspacing
G.~Jain, ``{CNN-LSTM model},'' \emph{Kaggle}, Feb. 2021, accessed: 2024-06-20. [Online]. Available: \url{https://www.kaggle.com/code/gaurav2022/cnn-lstm-95/notebook}
\BIBentrySTDinterwordspacing

\bibitem{Pervasive22}
E.~Baccour, et~al., ``Pervasive ai for iot applications: A survey on resource-efficient distributed artificial intelligence,'' \emph{IEEE Communications Surveys \& Tutorials}, vol.~24, no.~4, pp. 2366--2418, 2022.

\bibitem{awad2011energy}
A.~Awad, O.~A. Nasr, and M.~M. Khairy, ``Energy-aware routing for delay-sensitive applications over wireless multihop mesh networks,'' in \emph{2011 7th International Wireless Communications and Mobile Computing Conference}, 2011, pp. 1075--1080.

\end{thebibliography}
